\documentclass[a4paper,twocolumn]{emulateapj}
\slugcomment{}

\usepackage{epsfig,wrapfig,color,sidecap}
\usepackage{index}
\usepackage{subfigure}

\usepackage{colordvi}

\usepackage[usenames,dvipsnames]{xcolor}

\usepackage{times}
\usepackage[utf8x]{inputenc}
\usepackage{mathrsfs}

\shorttitle{}
\shortauthors{Benyamin et al.}

\RequirePackage[colorlinks=true
,urlcolor=blue
,anchorcolor=blue
,citecolor=blue
,filecolor=blue
,linkcolor=blue
,menucolor=blue
,pagecolor=blue
,linktocpage=true
,pdfproducer=medialab
]{hyperref}

\begin{document}

\bibliographystyle{authordate1}

\title{Recovering the observed B/C ratio in a dynamic spiral-armed cosmic ray model}

\author{David Benyamin$^1$,Ehud Nakar$^2$, Tsvi Piran$^1$ \& Nir J. Shaviv$^1$} 
\affil{1. The Racah Institute of physics, The Hebrew University of Jerusalem, Jerusalem 91904, Israel}
\affil{2. Raymond and Beverly Sackler School of Physics \& Astronomy, Tel Aviv University, Tel Aviv 69978, Israel}

\begin{abstract}

We develop a fully three dimensional  numerical code describing the diffusion of cosmic rays in the Milky Way.  It includes the nuclear spallation chain up to Oxygen, and allows the study of various cosmic ray properties, such as the CR age, grammage traversed, and the ratio between secondary and primary particles. This code enables us to explore  a model in which   a large fraction of the cosmic ray acceleration takes place in the vicinity of galactic spiral arms and that these spiral arms are dynamic.
We show that the effect of having dynamic spiral arms is to limit the age of cosmic rays at low energies. This is because at low energies the time since the last spiral arm passage governs the Cosmic Ray (CR) age, and not diffusion. Using the model, the observed spectral dependence of the secondary to primary ratio is recovered without requiring any further assumptions such as a galactic wind, re-acceleration or various assumptions on the diffusivity.  In particular, we obtain a secondary to primary ratio which increases with energy below about 1 GeV.

\end{abstract}


\section{introduction}

\maketitle

One of the most fundamental cosmic ray measurement relevant to the understanding of the cosmic ray propagation is  the ratio between the fluxes of primary source particles, and secondary particles formed as the particles propagate in the galactic halo .

Under the simplest leaky box or slab disk models, one expects the path lengths traversed by the cosmic rays to be inversely proportional to the diffusivity $D$. If $D \propto E^\delta$ (or $\propto E^{\delta/2}$ for non-relativistic particles), we expect higher energy particles to diffuse faster out of the galactic plane, implying that those particles that reach the solar system, do so earlier at higher energies \citep[e.g.,][]{CRreview}. As a consequence, higher energy particles should produce less secondaries along the way. Indeed, the ratio between the secondary Boron and primary Carbon nuclei decreases at high energies as expected \citep{GarciaMunoz,StrongReview,Dragon}.  However, at  lower energies, of order $E \lesssim 1$GeV$/n$ (GeV per nucleon), this behavior breaks down and  the B/C ratio increases with energy. 

One clue as to the nature of the models required to solve this discrepancy is found by looking at the ratio between a radioactive isotope, $^{10}$Be to a stable isotope $^{9}$Be, which measures the ``age" of the cosmic rays. Instead of increasing with energy at lower energies, the ratio has a plateau at $E \lesssim 1 GeV/n$. This implies that the cosmic ray age saturates at energies of $E \lesssim 1 GeV/n$ \citep[e.g.,][]{StrongReview}. This  can explain the apparently anomalous B/C behavior. This is because at low energies, the grammage traversed by the cosmic rays is proportional to a fixed age times the velocity, which increases with energy. Namely, the B/C ratio for a fixed age, increases  with energy, as observed.

Presently, there are three standard explanations (or a combinations of them) to the B/C anomaly, all of which give the same typical age for low energy particles \citep[e.g.,][]{StrongReview,Dragon}. 
First, one can assume that the diffusion coefficient is fixed below an energy of a few GeV$/n$. This is a rather ad hoc assumption, but it is possible if statistical properties of the ISM magnetic field change below a certain length scale.

A second type of models involves galactic winds. Since CRs cannot reside in the halo longer than it takes the wind to clear them out, the CR age at low energies approaches an energy independent value.  The winds themselves can be driven by the CR pressure. Finally, a third type of models includes reacceleration, whereby low energy particles are accelerated by the ISM itself. This can be described as diffusion in momentum space. As a consequence, CRs at different energies have similar ages. 

In principle, the different type of models all predict a similar B/C energy dependence, but there are several subtle  differences. For example, reacceleration should leave a signature on K-capture isotopes at low energies. Such a signature may be present, though it is yet inconclusive given the uncertainties in the nuclear cross-sections  \citep{Kcapture}.

In this paper, we examine a fourth type of solutions in which the CR age saturates at low energies because of CR source inhomogeneity and time dependance associated with the galactic spiral arms.  We show that this solution arises naturally  when inhomogeneity 
in the CR source distribution, namely in the distribution of SNRs within the milky way  is taken into account. 

Probably the least discussed hidden assumption of the CR diffusion models, which were until recently the state of the art \citep[e.g.,][]{StrongNucleons}, is that they generally consider the CR sources to be distributed in the galactic disk with cylindrical symmetry (that is, the distribution depends only on $r$ and $z$). Although it may be an adequate description for some CR characteristics, this does not have to be case. Cosmic rays from below 1GeV to much above $10^{15}$eV, are thought to
originate in supernova remnant (SNR)  shocks. This is supported by observations of synchrotron
\citep{Koyama} and inverse-Compton  \citep{Tanimori} emission of high
energy electrons in SNRs, as well as $\gamma$-ray emission, which probably
originates from high energy protons \citep{Aharonian}.
However, most SNe taking place in spiral galaxies like the Milky Way are core collapse SN \citep[e.g.,][]{vdBergh}. These arise  from massive stars that live a relatively short time (the lowest mass star which will undergo a core-collapse supernova lives about 30 Myr). 
This expected inhomogeneity of CR acceleration in spiral armed galaxies can be seen in  \cite{Lacey}, where one clearly observes in NGC 6946, that the majority SNRs are found in the vicinity of the spiral arms. Because star formation is not homogeneously distributed in the Milky Way, neither the SNR distribution nor the acceleration of CRs distributions can be expected to be homogeneous either.

Direct observational evidence for preferential acceleration in the Milky Way spiral arm includes a spectral breaks in synchrotron radiation and photons from $\pi^0$ decay. The $\gamma$-ray spectrum at 
300 to 700 MeV (corresponding to proton energies of about 7 to 20 GeV) shows that the cosmic ray spectral index in the direction of the galactic arms is harder by 0.4$\pm$0.2 than the index in other directions \citep{rogers,Bloemen}. Similarly, the electron synchrotron spectrum at 22-30 GHz, produced by cosmic ray electrons at around 30 GeV is harder in the direction of the galactic arms than in other directions \citep{Bennett}. Both observations cannot be explained from an azimuthally symmetric source distribution. The synchrotron requires the sources to be concentrated around the arms, such that further away synchrotron and inverse-Compton cooling will soften the electron spectrum. The $\pi^0$ decay spectrum requires even more. Since it cannot be explained with just diffusion from a time-independent source distribution, one of the basic assumptions, such as time independence, must break. 
 
The source inhomogeneity was also shown to be important when one considers the long term temporal variations in the CR flux reaching the solar system \citep{ShavivNewAstronomy}. In this work, it was shown using the CR exposure age of Iron meteorites that the CR flux has been variable over the past 10$^9$yr due to passages through the galactic spiral arms. It was shown that the CR flux varies by at least a factor of 2.5 between when the solar system is between the spiral arms, and when it is inside the arms.  
Clearly then, an obvious modification to the galactic diffusion models would be to consider taking the CR sources to be distributed unevenly, with concentration in the spiral arms. 

Inhomogeneity and in particular source concentration in galactic arms has the potential to resolve the interesting inconsistency between the standard diffusion models and observation in  the so called ``Pamela Anomaly". The {\sc pamela} satellite was used to measure the $e^+/(e^++e^-)$ ratio in CRs at different energies. Since Positrons are secondary particles, the ratio is expected to decrease with energy. However, the observations revealed that the ratio increases with energy, from about 10 GeV to 100~GeV (above which there is insufficient statistics, \citealt{Adriani}). Since standard diffusion could not explain this behavior, it was  immediately assumed that the increased Positron population is part of a hard spectrum of pairs (harder than the normal cosmic ray population) such that it dominates the normal secondary positrons above 10 GeV. This can arise  if this pair population is part of the annihilation spectrum of dark matter particles \citep{DarkMetter1,DarkMetter2}. Astrophysical explanations in which pairs are directly accelerated in different astrophysics sources (particularly in pulsars, \citealt{Pulsar1}, \citealt{Pulsar2}, \citealt{Pulsar3}, \citealt{Pulsar4}, and \citealt{Pulsar5}) or in aged SNRs (\citealt{positrons-secondaries1}, \citealt{positrons-secondaries2}, \citealt{positrons-secondaries3}, and \citealt{positrons-secondaries4} have also been proposed. 

Unlike these explanations, where a pair population is not produced through interaction with the ISM, \cite{Pamela} have shown that by considering a SNR distribution that is  primarily located near the galactic spiral arms, one naturally finds that the Positron fraction should start increasing above 10~GeV. To understand this behavior, one has to consider that Electrons and Positrons cool through synchrotron radiation (from the galactic magnetic field) and inverse-Compton scattering of the IR and visible light. This implies that if a source of electrons is located at a given distance from the solar system, only Electrons with a low enough energy can reach the solar system before cooling. Positons, however, are formed from Hadronic cascades, from Protons which can reach the solar vicinity even if they were formed far away. Thus, above the energy for which Electrons cool before reaching the solar system, the Positron to Electron ratio will increase. This fraction should increase up to at most a few 100~GeV until the local (disk) population of ``fresh" electrons takes over. 

Clearly then, source inhomogeneity should be taken into account as it has been in recent models \citep{ShavivNewAstronomy,Pamela,DRAGONspiral}. In the present work we show that an additional component is necessary, which is that the source distribution is {\em dynamic}. We show below that with this additional component one can recover the low energy behavior of the B/C ratio, without requiring a galactic wind, re-acceleration or various assumptions on the diffusivity.
In order to do so  we have developed a Monte Carlo code that  simulates the propagation of CRs in the dynamic galactic halo, taking into account both the inhomogeneous distribution and the motion of the galactic arms relative to the galaxy (and to the sun) . 
Note that it is not our goal here to carry out a full fledged parameter study. Instead, we focus on demonstrating the existence of the effect and its consistency with a full realistic numerical simulation. 

We begin in \S\ref{sec:analytical} by qualitatively estimating the effect expected from sources distributed preferably in the spiral arms and why it has the potential for predicting the observed secondary to primary ratio. In \S\ref{sec:model}, we describe the Monte Carlo code for general inhomogeneous source distribution that we developed.  We describe in \S\ref{sec:BC} the results for the B/C ratio, and their  implications to the values of the CR diffusivity and halo size. The implications of these results are summarized in \S\ref{sec:discussion}. 

\section{Understanding the effect of spiral arms}
\label{sec:analytical}
Before solving the full model numerically, we explore qualitatively the expected change in behavior once we introduce the spatial inhomogeneity and arm dynamics. To do so we examine, in this section,  a very simple approximation, in which all CRs come from the nearest spiral arm at a distance $s$ from us.

\subsection{The secondary to primary spectrum}
\label{sec:BoCana}

There are three important time scales describing this system\footnote{At this point, we neglect ``cooling" by Coulomb scattering which becomes important at low energies. It will be addressed in the numerical model.}. The first is the typical time it takes CRs to diffuse out of the galactic halo:
\begin{equation}
\tau_{escape}\approx {Z_h^2 \over 2 D},
\end{equation} 
 where $Z_h$ is the half width of the galactic halo and $D$ is the diffusion coefficient. The second time scale is the typical time it takes CRs to diffuse from the spiral arm to the solar system ($s$ is the distance from the solar system to the nearest spiral arm):
 \begin{equation}
 \tau_{s}\approx {s^2 \over 2D}.
 \end{equation}

The two time scales scale with energy through $D^{-1}$ and therefore their ratio is energy independent. That is, if some fraction of the CRs are lost from the galactic halo before they reach the solar system, it will be the same fraction at all energies. 

Which time scale dominates depends on the question asked. For example, if we care about the average grammage of a steady source originating from spiral arms, then the average grammage will be proportional to $\max(\tau_{escape}, (\tau_{escape} \tau_s)^{1/2})$ \citep{Pamela}. Namely, the effective diffusion time scale is the larger of the diffusion out of the galactic halo or the geometric mean of the two diffusion time scales.  

The third time scale is the time since the last passage of the solar system through a spiral arm\footnote{Note that there is a 15 Myr lag between the actual spiral arm passage and the average SN wave \citep{ShavivNewAstronomy}, we shall assume henceforth, that the spiral arms passage is defined by the average SN wave.}: 
\begin{equation}
\tau_{arm} \approx {s/ \sin i \over v_{arm}},
\end{equation}
where $v_{arm}$ is the pattern speed of the arm relative to the local ISM, $i$ is the pitch angle of the arm\footnote{We shall assume for simplicity that it also corresponds to the nearest arm. It need not be the case if the solar system is already near the next arm.},  and $s/\sin i$ is approximately the distance that the solar system has traversed from the last arm passage.

At sufficiently low energies, $\tau_{arm}$ is the shortest time scale, and it therefore dominates. Hence, if it takes say 20 Myr for cosmic rays at a given energy to diffuse from the spiral arms to the solar system, but the solar system last passage through the spiral arms was 10 Myrs ago, then the CRs in the solar vicinity should be around 10 Myrs old, rather than  20 Myrs (unless the diffusion out of the  galactic halo is significantly longer than the diffusion from the spiral arm).

Given the typical age $t$ of the CRs, which may be determined by diffusion from the arm, or out of the galactic halo, or by advection from the spiral arm, we can estimate the amount of secondaries produced as a function of energy. 
Over a typical time $t$, the grammage $g$ seen by a propagating CR is:
\begin{equation}
g \approx x \bar{\rho} \approx t v \bar{\rho},
\end{equation}
where $x = t v$ is the physical path length travelled by the CR particle moving at a speed $v$ and $\bar{\rho}$ is the average density along the path. 

There are four different regimes for the solution depending on whether the propagation is dominated by  advection or by diffusion  and whether the particle is relativistic or not.

In the limit dominated by  diffusion one has $t \sim \max (\tau_{escape},(\tau_{escape} \tau_s)^{1/2}) = 
\max (Z_h^2,s^2)/2D$. 
The dependence of the age on energy depends, in turn, on the  energy dependence of the diffusivity.  
We assume that the effective mean free path of the cosmic rays depends on the rigidity as $\ell \propto R^\delta$. For Kolmogorov turbulence, we should expect $\delta \sim 1/3$, but any value up to 0.6 is theoretically feasible. The diffusivity is then of order $D \sim \ell v$. 
For relativistic particles, $R \propto p \propto E$, and therefore:
\begin{equation}
g_\mathrm{rel,diff} \propto {\max (Z_h^2,s^2) \over E^\delta }  \bar{\rho} .
\end{equation}
For non relativistic particles, $R \propto p \propto \sqrt{E_{kin}}$, and:
\begin{equation}
g_\mathrm{non-rel,diff} \propto {\max (Z_h^2,s Z_h) \over E^{\delta/2} }  \bar{\rho}. 
\end{equation}

At lower energies,  in the advection dominated regime $t \sim \tau_{arm} \approx s / (v_{arm} \sin i) $. This life time  is independent of energy. Therefore:
\begin{eqnarray}
g_\mathrm{rel,adv} &\propto & {sc \over {v_{arm} \sin i}} \bar{\rho} \propto {\mathrm const.},\\ 
g_\mathrm{non-rel,adv} &\propto & {sv\over {v_{arm} \sin i}} \bar{\rho} \propto E^{1/2}. 
\end{eqnarray}
In particular, in the advection dominated non-relativistic case, the grammage is an increasing function of energy. This is one of the cardinal results of this work. 

\begin{figure}
\centerline{\includegraphics[width=3.4in]{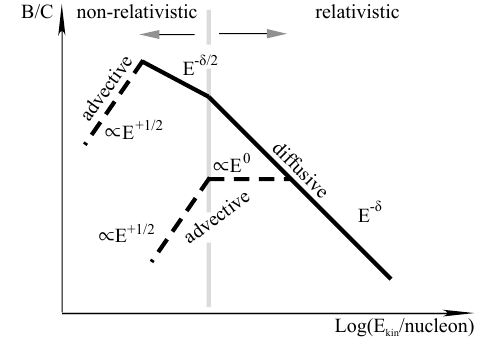}}
\caption{A heuristic description of the expected local interstellar B/C ratio. The solid lines describe the limit in which CR diffusion dominates the age of the cosmic rays, while the dashed lines denote regions where advection from the spiral arm dominates the CR age. The two behaviors depend on whether the advection/diffusion change of behavior takes place for relativistic or non-relativistic particles. At low energies (below a few 100 Mev/n, this picture is modified due to Coulomb collisions in the ISM, and solar wind modulation).}
\label{fig:heuristicBC}
\end{figure}

If we combine the results together, we find that there are two net possible energy dependences, depending on whether the diffusion/advection behavior change takes place for non-relativistic or for relativistic particles. The results are summarized in fig.\ \ref{fig:heuristicBC}.

 Under standard diffusion models the velocity cancels out (e.g., see \citealt{StrongReview}, p. 295), such that to obtain an increase in the B/C ratio one requires a non-standard diffusion model, having a diffusion coefficient that depend not only on the rigidity, (\citealt{DRAGONnon-standard}).  These results should be compared with the standard solutions to the B/C behavior.  The  B/C behavior at low energies requires a grammage which increases with energy. Therefore any solution for which the time it takes the CRs to reach the solar system saturates at a finite value that is independent of the energy can recover the B/C behavior. One option is simply to assume that the diffusivity below $E_{crit}/\mathrm{nucleon} \sim 4$GeV is constant. 
Here the age saturates at $\tau_{max} \sim Z_h^2/D(<E_{crit})$. In the solution where a galactic wind with a velocity $v_{wind}$ is added the age saturates at $\tau_{max} \sim Z_h/v_{wind}$, the time it takes the wind to evacuate the galactic halo from CRs. Another standard solution is that of re-acceleration. Here low energy particles are mixed giving them the same average age.

The break energy between the rising B/C and the decreasing B/C can be obtained by comparing the relevant diffusion time scale to the advection time scale, $\max(\tau_{escape},\tau_s) \approx \tau_{arm}$. Assuming that the break takes place when the particles are relativistic we have:
\begin{equation}
E_{break} \approx E_0 \left( s v_{arm} \sin i \over 2 D_0 \right)^{1/\delta},
\end{equation} 
where we assumed a diffusion coefficient of $D = D_0 (E/E_0)^\delta$.

For example, if we reasonably assume that the galaxy has 4 arms with a 28$^\circ$ pitch angle, rotating with a periodicity (arm passage relative to the solar system) of 145 Myr, and a diffusion coefficient $D = D_0 (E/E_0)^\delta$ with $D_0 = 2 \times 10^{27} cm^2/s$, $E_0 = 3$ GeV and $\delta = 0.5$, we find that $E_{break} \approx 1.5$ GeV if the last spiral arm passage was 5 Myrs ago. This break energy  will be around 1GeV and can explain the observed B/C behavior. However, a full numerical calculation is needed to determine the accurate spectrum and the break obtained from a given set of parameters. Moreover, one has to remember that additional factors come into play at this energy range,  such as solar modulation and Coulomb cooling. 

\section{The numerical model}
\label{sec:model}

The model we develop here includes many of the standard constituents found in models, i.e., diffusion in a galactic halo with an energy dependent diffusion coefficient, and nuclear interactions with the ISM (which can produce secondary particles). We do not include galactic winds, re-acceleration or ad-hoc assumptions on the diffusivity,  however we do include dynamic spiral arms. That is, the source distribution is inhomogeneous and non-axisymmetric and time dependent. As we have seen in \S\ref{sec:analytical}, this time dependence is important in determining the  the secondary to primary ratio.

The present numerical model  is a full 3D extension to the ``inhomogeneous" 2D models of \cite{ShavivNewAstronomy} and \cite{Pamela}. The two previous models assumed that the spiral arms are straight rods and that all the physics is in the plane perpendicular to these rods. Here, we relax this assumption and build a full 3D model for the Milky Way with a dynamic spiral arms distribution. 

\subsection{The Source Distribution}
\label{sec:sources}

The source of cosmic rays at the relevant energy range is supernova remnants. We divide the supernova population into degenerate collapse SNe (type Ia) and core collapse SNe (type II and Ibc). Given the observational data of SNe in different spiral armed galaxies, we expect the type Ia contribution to be around 15-20\% \citep{Tammann,vdBergh}. Here we take 20\%. 

Since the progenitors of core collapse SNe live less than 30 Myr, they explode not far from their birth sites, which is  primarily the spiral arms of the galaxy \cite[e.g., see][]{Lacey}. Because about 5\% of O stars and 15\% of HII regions reside outside the spiral arms (see \citealt{ShavivNewAstronomy} and refs.\ within), we assume here that 10\% of the core collapse SNe are in the ``field", i.e., distributed in a homogeneous disk, while the rest divided equally between 2 sets of arms (see \S\ref{sec:arms}). We note that this disk is different and generally narrower than the disk-like halo in which CR diffuse.  

For the actual radial and vertical distribution of Type Ia and core collapse SNe, we take the respective distributions in \cite{Ferri}. For the sets of arms, we take the radial and vertical distribution of the field SNe and add an azimuthal component, as in \cite{ShavivNewAstronomy}, with spiral arm location and dynamics described below. 

\subsection{The Spiral Arms}
\label{sec:arms}

There are several indications that the Milky Way has two sets of spiral arms \cite[e.g., see][and references therein]{AP1997,Naoz} one with an $m=4$ symmetry, and one with an $m=2$ symmetry. This can also be seen from the work of \cite{Gavish}, in which the two different sets can be from $ln(r) - \phi$ plots based on the CO maps of \cite{Dame}.  The nominal numbers we take  are $i=28^\circ$ for the four-armed set, and $i=11^\circ$ for the two-armed set. 

We approximate the spiral arms  as logarithmic spiral arms, as is expected from the density wave theory \citep[e.g.,][]{BT}:
\begin{equation}
r_j = a_j \exp \left[ b_j (\phi - \phi_{0,j} )\right],
\end{equation}
where $a_j$ is the inner radial extent of the arm $j$, and  $b_j={\tan \left(  i_j \right)}$ where $i_j$ is the pitch angle of arm $j$ and $r$, $\phi$, and $z$ are cylindrical coordinates centered on the milky way galaxy.

Given that there are two sets of spiral arms, it is not surprising that different groups have found different pattern speeds which cluster around two values (see the summary in \citealt{ShavivNewAstronomy}). This is also why an unbiased analysis of open clusters, which allows for multiple patterns, found both pattern speeds \citep{Naoz}.  
Following the latter analysis, we take $\Omega_2= 25 (km/s)/kpc$ for the two armed set (which is nearly co-rotating with the solar system, see below), while $\Omega_4= 15 (km/s)/kpc$ for the 4-armed set. 

We take the rotation curve of \cite{Olling} in our model.

\subsection{The Diffusion Coefficient}

We assume that the cosmic rays undergo normal diffusion without additional advective processes, relative to the interstellar medium which itself is differentially rotating around the galaxy. This diffusion coefficient is assumed to be location independent up to the galactic halo height $Z_h$, beyond which CRs can escape without diffusion. 

The diffusion coefficient is also assumed to be energy dependent, with the general form $D = D_0 \beta R_{GV}^\delta$, where $\beta = v/c$ and $R_{GV}$ is the rigidity  in GV.  $\delta$ is the diffusion coefficient power-law index, which depends on the physical model for the diffusion. 

One  can show that a Kolmogorov power-law spectrum for the interstellar turbulence having $\delta B \approx 5 \mu G$ gives rise to a diffusion coefficient $\kappa \sim 2 \times 10^{27} \beta R_{GV}^{1/3} cm^2/s$ \cite[see][and references therein]{StrongReview}, i.e., $\delta = 1/3$. Other, Non-Kolmogorov models for the ISM magnetic turbulence can result different power laws. For example, Kraichnan-type turbulence produces a steeper spectrum, with $\delta=1/2$ \citep{ref56}.

Therefore, CR diffusion models typically consider a diffusion power law index of $\delta \sim 0.3$ to $0.6$ \citep{StrongReview}. We use a nominal value of $\delta = 1/2$ as it appears to best explain the high energy dependence of B/C. 
  
\subsection{Nuclear Cross-Sections}
\label{sec:NucX}

A major ingredient of the model is the nuclear spallation reactions which the CRs undergo during their propagation. Since we here concentrate on light isotopes, we consider all the spallation and decay reactions of isotopes up to Oxygen. Given that the contribution by the heavier nuclei, between Fluorine and Silicon, is only about 11\% of the total mass between Carbon and Silicon, one would at most get a 10\% correction if all the above-Oxygen elements break up to Boron and Carbon. We therefore neglect the contribution of the heavier elements in this work, and keep in mind that subsequent analyses will require including these elements if a few percent level accuracy is needed.  A summary of the reactions we consider can be found in \cite{GarciaMunoz}.  

Since the typical time scale for diffusion and for spallation is 10$^7$yr, there is no need to evolve all the short radioactive isotopes 
separately. That is, when a short lived isotope is formed, it is immediately converted into its daughter products. If there is a branching ratio between different decay processes, then each decay branch is randomly followed with the proper weight. Below oxygen, there is only one long lived radioactive isotope which is explicitly followed, and that is $^{10}$Be. 

For the partial spallation cross-section, we use the {\sc yieldx} program of \cite{Xsection}. We use the formulae of \cite{Karol1975} for the total inelastic cross-sections of interactions with helium.

We assume a fixed ratio between Oxygen and Carbon at the source. As we shall see in \S\ref{sec:CO}, this ratio, of 1.45, is obtained by fitting to the observed C/O ratio.

\subsection{Coulomb and Ionization Cooling}

At energies below about 1GeV, the propagating nucleons lose energy through primarily Coulomb collisions and ionization loses. The expression for Coulomb collisions is approximately  \cite[e.g.,][]{StrongNucleons}
\begin{equation}
\left. dE \over dt \right|_{\mathrm{Coul}} \approx -{4 \pi r_e^2 c^3 m_e Z^2 n_e \ln \Lambda \over \beta},
\label{eq:CoulCool}
\end{equation} 
where $\ln \Lambda \sim 40-50$ is the Coulomb logarithm and $r_e$ is the classical electron radius. This expression assumes that the typical velocity $\beta c$ is much higher than the thermal speed of electrons in the ISM.

Eq.\ \ref{eq:CoulCool} can be written as an energy loss with grammage, once we consider that $dg \approx {(n_p m_p \beta c / X)} dt$. Here $X$ is the Hydrogen mass fraction.
\begin{equation}
\left. d(E/A) \over dg \right|_{\mathrm{Coul}} \approx -{4 \pi X \left(n_e \over n_p\right)  \left(m_e \over m_p\right) \left(Z^2 \over A \right)  {r_e^2 c^2 \over \beta^2} \ln \Lambda}.
\end{equation} 
When written in this form, it is clear that for each specie, the Coulomb ``cooling" is proportional only to $\beta^{-2}$. This expression can be easily incorporated into the simulation, as described below.

The ionization loses have a similar expression to the Coulomb loses. The difference is that instead of $\ln \Lambda$ there is a more complicated expression that we use \cite[see][]{StrongNucleons}. For typical ISM conditions, the ionization losses below 1 GeV are roughly 1/5 of the Coulomb collisional losses. 

\subsection{Solar Modulation}
\label{sec:solmod}
The cosmic ray spectrum observed at Earth is different from the spectrum outside the solar system because of interaction with the solar wind. To a good approximation, this interaction can be quantified through the solar modulation parameter $\phi$, which is the energy per proton lost by the CR particle \cite[e.g.][]{UsoskinPhi}. The energy per nucleon lost is
\begin{equation}
 (E/A)_{earth} = (E/A)_{ISM} - (Z/A) \phi .
\end{equation}  
Typically, $\phi$ ranges between 300 MV to 500 MV in the periods when the measurements are taken \citep[e.g.,][]{UsoskinPhi}, which is the range we take here. 

\subsection{Diffusion with arms}
\label{sec:diffusion_with_arms}

The CRs spatial distribution from the axisymmetric components does not depend on the rotation of the underlying interstellar medium. However, when considering the diffusion from the spiral arms, the rotation of the arms and the rotation of the ISM  should be considered. 

We solve the diffusion from each spiral set (as well as from the galactic disk) separately. For each set, we solve in the frame of reference of the moving spiral arm set. On one hand, it implies that we need to add the differential rotation of the galactic disk as a radially dependent drift velocity, $V_{arm}(r)-V_{disc}(r)$, which may appear as a complication. On the other hand, it turns the time dependent problem, into a time independent one. This is because the boundary conditions in the frame of reference which is rotating with the spiral arm are fixed in time. 

Thus, at a given time, the CR density distribution is obtained by rotating the distribution from each set by the appropriate rotational phase and adding the distribution from the axisymmetric sources. 

\subsection{Initial Composition}
\label{sec:initial_composition}
For the initial relative abundances for the primary elements, we take 40\% Carbon, 2\% Nitrogen and 58\% Oxygen. The relative abundances of each isotope for a given element are assumed to be solar.  As we shall see in \S\ref{sec:CO}, these abundances recover the observed C/O and N/O ratios.

\subsection{ISM distribution}
\label{sec:ISM_distr}
For the ISM components we take the distributions given in \cite{Ferri}. The total ISM density distribution is required for the determination of various CR/ISM interactions.

\subsection{The numerical algorithm}

Our code is a Monte Carlo simulation. This implies that we solve the full diffusive propagation of a large sample of particles, having different initial energies, different origins (arms or disk) and representing different compositions. The particles are followed from their formation until they either leave the galaxy, break into lighter particles than Beryllium or cool below 100 MeV. For each particle, we also follow its various interactions, including reactions which change their identity, that is, spallation on the ISM and radioactive decay. When this occurs, we follow the lighter particle that formed. By simulating many particles, with different initial energies, we can construct the CR spectrum at different locations in the galaxy, and at different times. 

The methodology we chose to follow, that of a Monte Carlo simulation, is different from most present day simulations (such as {\sc galprop}, \citealt{StrongNucleons} and {\sc dragon}, \citealt{Dragon}), which solve the diffusion PDEs. The main advance of the latter type is that the accuracy increases linearly with the number of grid points. The disadvantage is that it is much harder to include new physical ingredients, such as time dependence or different geometries. 

The price one pays when simulating using the Monte Carlo method is that the accuracy increases only as the square root of the number of particles used. However, there are several major advantages. First, it is almost straightforward to include any physical effect. In our case, we require to have spatially and temporally dependent CR sources associated with the spiral arms. The code is also easily parallelizable because each simulated CR particle is independent of other CRs.

Another very important advantage, is that the different CR variables, including artificial ones, can be easily recorded. The best example is the CR age and the amount of grammage that the CR has traversed. This allows obtaining the path length and age distributions. In fact, if one artificially switches off cooling, it is possible to see at low energies multiple passages of the spiral arm in the PLD. This is impossible to obtain when solving the diffusion PDEs. 

We have used  $10^{10}$ CR particles in the simulations. For each CR, we randomly choose an initial isotopic identity, as described in \S\ref{sec:initial_composition}.
We then continue to randomly choose the initial location in the galaxy, where each particle was accelerated, with a distribution described in \S\ref{sec:sources}.

Instead of simulating the small steps corresponding to the physical mean free path (m.f.p.) \ of the cosmic rays, around 1~pc, our MC simulation uses a larger {\em effective} m.f.p., which lumps together many small effective physical steps. This corresponds to a larger effective time step chosen to be $\tau_\mathrm{esc}/100$. For our nominal model, this corresponds to an effective m.f.p.\ of 25~pc. To avoid numerical error, we keep the effective m.f.p.\ smaller than the typical length scale over which the gas density varies. To maintain accuracy, we decrease the time step a 1000-fold when the CR particle is in the vicinity of the solar system (within a sphere of radius 0.2 kpc). This roughly corresponds to an effective m.f.p.\ of 1~pc, which is of order of the real m.f.p.\ .  Each time step, we check whether the particle had left the galaxy, and if it did not, we calculate, using the cross-sections described in \S\ref{sec:NucX},  the grammage that the particle traversed in this time step. We then calculate the probability that the particle had undergone a nuclear reaction with ISM nuclei. If the particle did break into a secondary particle, we then follow each secondary particle with the same methodology until it either breaks into a particle lighter than Beryllium, until it escapes the galaxy or until it cool below {{100~MeV.}}

Although the energy and location of each simulated particle is continuous, both are recorded with a set of 4 dimensional arrays (location + energy), for each of the stable isotopes between $^9$Be to $^{16}$O, and the long lived $^{10}$Be.

The CR particles are not restricted to a grid, but the densities are recorded on a coarse grid covering the galaxy, and a finer grid within a sphere of radius 0.1 kpc around the solar system. The coarse grid spans a volume of 30kpc $\times$ 30kpc $\times$ 2$Z_h$, where $Z_h$ is the height of the galactic halo, which depends on the chosen model parameters. It is composed of (0.1kpc)$^3$ sized cells. Within the finger grid includes (1pc)$^3$ sized cells. The energies are recorded in 10 bins per decade, covering the range of 0.1-1000 GeV.

\section{Results}
\label{sec:BC}

Table 1 summaries the three models presented here. Model A is a simple disk model having standard diffusion parameters but having the spallation cross-sections fixed at their 1GeV value and with the cooling switched off. This stripped model can be used to demonstrate the basic energy dependence of the secondary to primary ratio, and the effects of saturation at large ``optical depths". Model B is a full homogeneous model. It is  used to test our code by running a case similar to that found in the literature \citep{StrongNucleons}.  The results of  both Model A and Model B are described in \S\ref{sec:modelsAB}.

Model C is our inhomogeneous model with a dynamic spiral arms distribution which bests fits the observed data on secondary and primary isotopes all the way up to Oxygen. As we elaborate in the discussion, the path length distribution in the present model is not exponential as is the case in the leaky box model, or nearly the case in the homogeneous disk model. Consequently, some of the basic diffusion parameters are necessarily different, in particular, the halo size and diffusion coefficient. For this reason, we must vary these parameters to recover the overall secondary to primary ratio and the $^{10}$Be/$^{9}$Be ratio. Since the rise with energy of the secondary to primary ratio depends in our model on the time since the last spiral arm passage, the third parameter which was varied is the time since the last spiral arm passage. 

The power-law index of the diffusivity, $\delta$ is of particular theoretical interest. We choose to work with $\delta = 1/2$ here, as it will best recover the B/C ratio. However, we note that a lower $\delta$ will provide a better fit to the C/O and N/O ratios as described below in \S\ref{sec:CO} (and to the positron fraction described elsewhere, \citealt{Pamela}). Namely, there is an inconsistently between the energy dependence of the different nuclear ratios which we do not understand and cannot resolve at this point. 

The other parameters describing the model, such as the ISM gas and source distributions and the geometry of the arms are all kept fixed at our nominal values described in \S\ref{sec:model}. This is done for practical reasons. Since our numerical model is CPU intensive, it is hard at this point to study the whole parameter-space.  

\subsection{Comparison to Homogeneous Models}
\label{sec:modelsAB}

\begin{table*}
\centering
\caption{Model summary}
\begin{tabular}{ c c c c c c c c}
\hline
 & model description & $D_0$ & $Z_h$ & $\delta$  & arms & cooling & cross-sections\\
\hline
Model A & ``stripped" disk model &$7 \times 10^{27} cm^2 /s$  & $1kpc$ & $0.6$ & - & - & fixed\\
Model B & {\sc galprop}-like disk model (no extra physics) & $7 \times 10^{27} cm^2 /s$ & $1kpc$ & $0.6$ & - & + & real\\
Model C & nominal model with arms & $5.5 \times 10^{26} cm^2 /s$  & $250pc$ & $0.5$ & + & + & real\\
\hline
\end{tabular}
\end{table*}

\begin{figure}[ht]
\centerline{\includegraphics[width=3.0in]{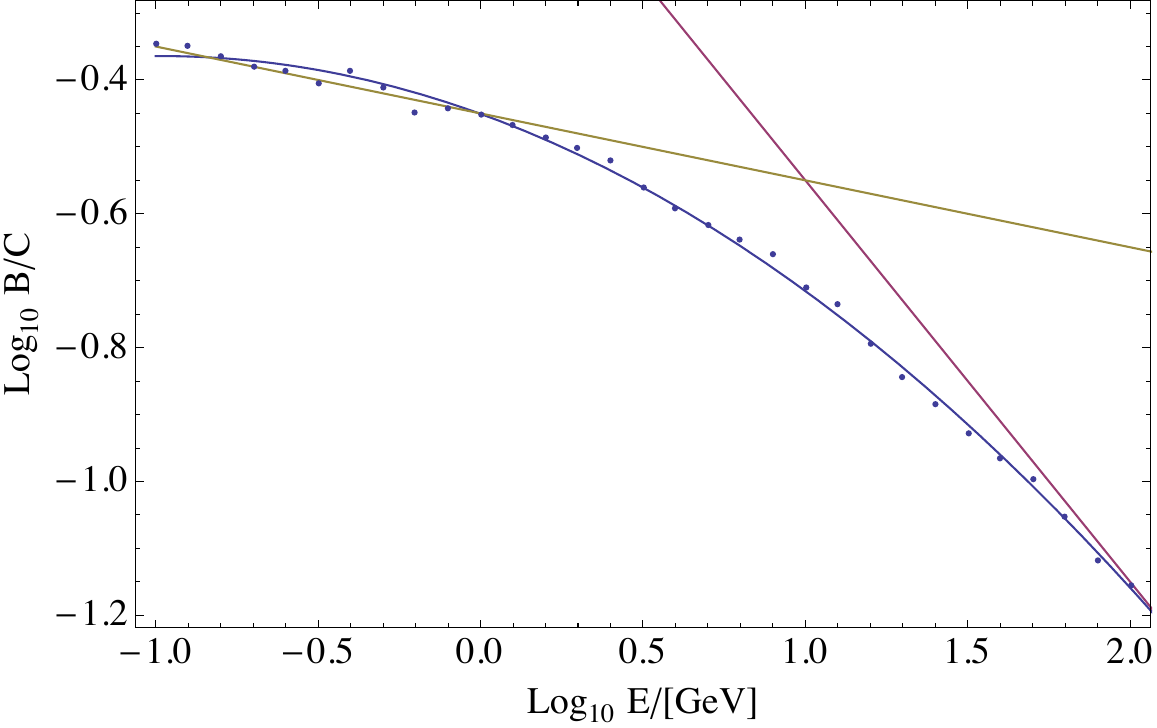}}
\caption{The B/C ratio obtained in Model A---a homogeneous disk model with $D_0=7 \times 10^{27} cm^2 /s$, $Z_h=1kpc$ and $\delta=0.6$, while imposing cross-sections which are fixed at their 1GeV values. The power law we obtain should be compared with the analytical estimates given in \S\ref{sec:BoCana}. Note that above 10 GeV, the slope approaches the value of $\delta=0.6$ (marked with the dark red line). At lower energies, the slopes are shallower than the analytical estimates because of saturation (the optical depth for spallation becomes of order unity). Between 0.1 and 1 GeV, for example, the slope (denoted by the mustard colored line) is about 0.1 instead of 0.6/2 of the optically thin non-relativistic limit. }
\label{fig:FixXsection}
\end{figure}

\begin{figure}[h]
\centerline{\includegraphics[width=3.0in]{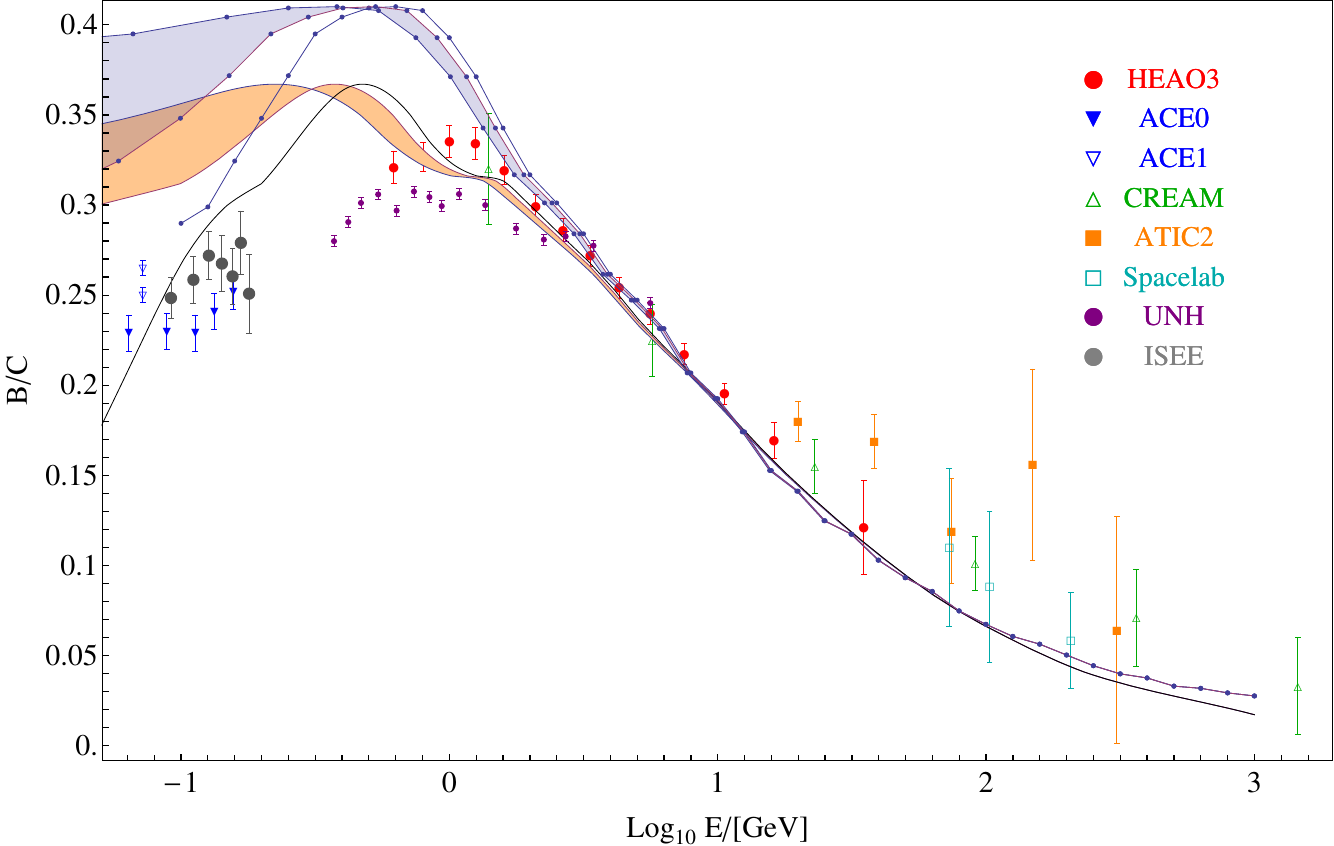} }
\caption{B/C for in Model B (dotted line and blue shading) having the same parameters as model {\sc galprop01000} of \cite{StrongNucleons} (smooth black line and orange shading), namely, a homogeneous disk, $D_0=7 \times 10^{27} cm^2/s$, $Z_h=1 kpc$, $\delta = 0.6$ and no additional physics such as a galactic wind or reacceleration. The shaded regions correspond to the spectrum once solar wind modulation is added (see \S\ref{sec:solmod}). Model B roughly recovers the {\sc galprop} results, demonstrating the general validity of our model   (see \citealt{StrongNucleons}).}
\label{fig:01000}
\end{figure}

\begin{figure}[h]
\centerline{\includegraphics[width=3.0in]{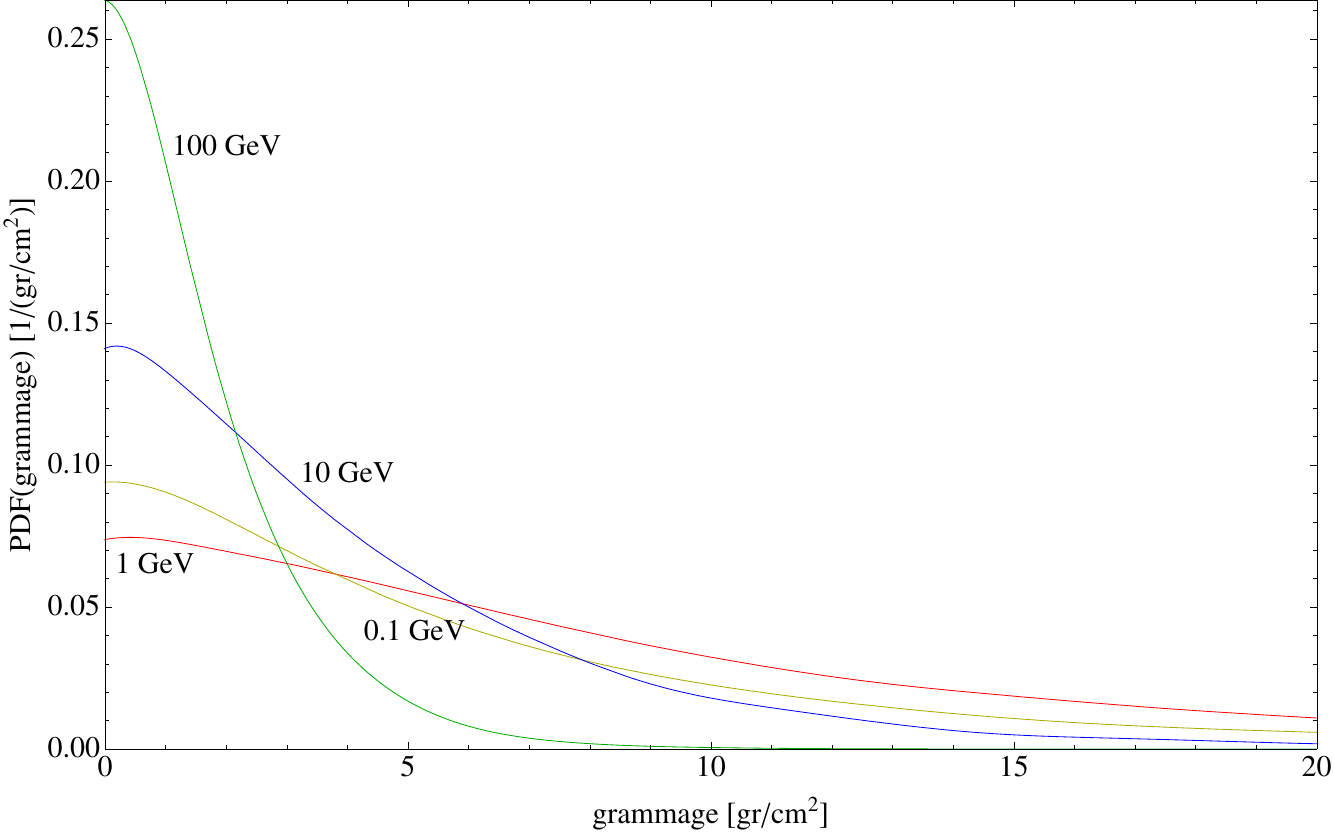}} 
\caption{The Path Length Distribution in Model B, with $D = 7 \times 10^{27}cm^2/s$. The PLD distributions are close to an exponent.  }

\label{fig:PLDgalprop}
\end{figure}

We begin with ``model A", by comparing our results with the simplest homogeneous model, to verify the validity of the numerical methods.

We first simulate the B/C ratio obtained in a model for which the spallation cross-sections are artificially fixed to their values at 1GeV/nucleon. This allows us to verify that the energy dependences are the power-laws we expect (see \S\ref{sec:BoCana}). fig.\ \ref{fig:FixXsection} demonstrates that the Monte Carlo simulation agrees well with the analytical approximation. Specifically, we obtain the expected broken power-law driven by the diffusion law, with a break at 1GeV/nucleon associated with the change from the non-relativistic to relativistic regime. At non-relativistic velocities, the power law is still less than $\delta/2$ because of the saturation effect of the spallation.

The next model we simulate, Model B, includes the energy dependent nuclear cross-sections and cooling, such that that we recover the results of {\sc galprop}'s simplest model (specifically, model 01000 of \citealt{StrongNucleons}, having no break in the diffusion-law, no wind and no re-acceleration). The model has the following parameters: $D_0=7 \times 10^{27} cm^2/s$, $Z_h=1kpc$ and $\delta=0.6$. $D_0$ is normalized at a rigidity of 3 GeV/nucleon. The results are plotted in fig.\ \ref{fig:01000}, which should be compared with fig.\ \ref{fig:01000} of \cite{StrongNucleons}.  Although there are several small differences the general agreement is apparent. The inconsistency at low energies can be explained given that the two models still have small differences. These include different cross-section tables as well as different spatial distributions for the SNe and the ISM density.  
Unlike the {\sc galprop} \citep{StrongNucleons} and {\sc dragon} \citep{Dragon} simulations, our present simulations can be used to obtain the path length distributions. These are plotted in fig.\ \ref{fig:PLDgalprop}. As expected, the distributions are exponentials. This also explains why the leaky box model is generally a good approximation for CR diffusion in homogeneous disk.   

\subsection{General results with spiral arm dynamics}

\begin{figure*}[ht]
\centerline{\includegraphics[width=3.0in]{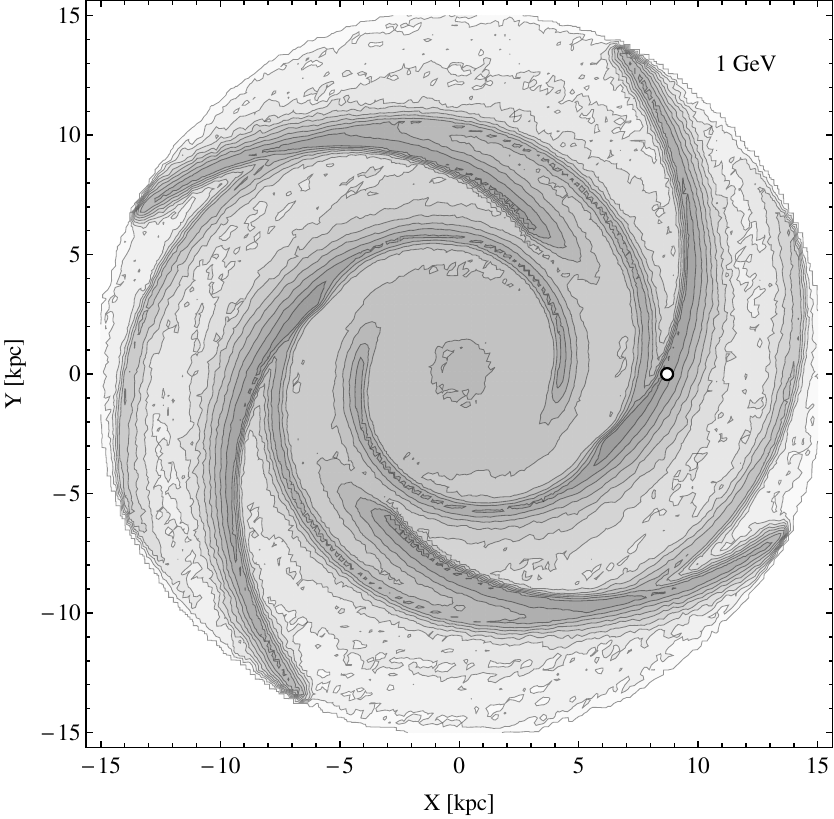} 
\includegraphics[width=3.0in]{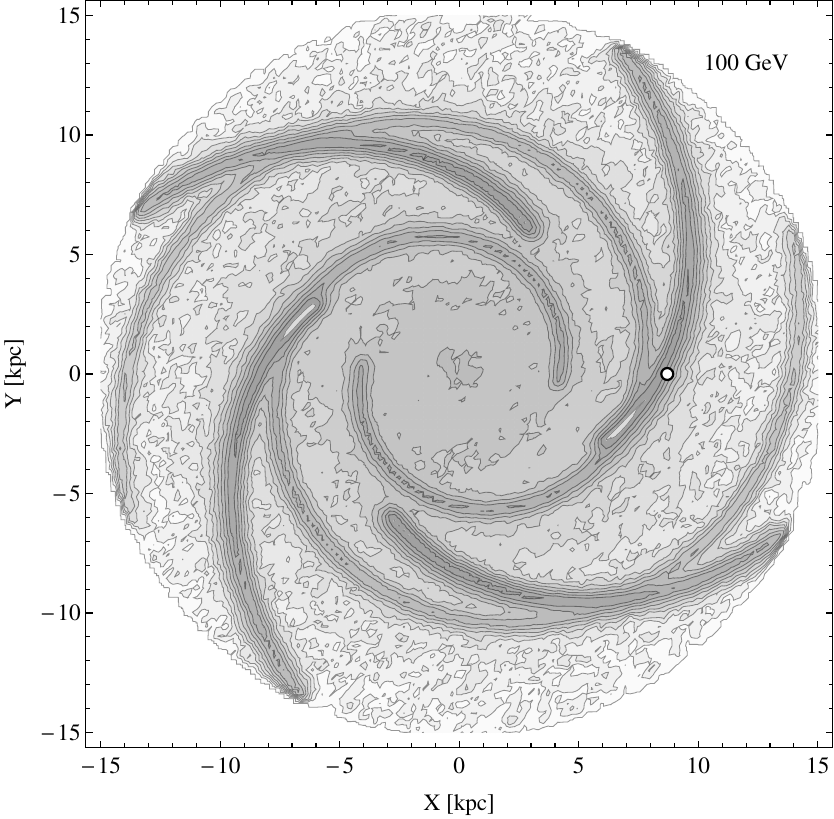}}
\caption{The relative CR density distribution in the Milky Way at two energies, normalized to the density at the location of the solar system. Left: At 1 GeV, Right: At 100 GeV. The contours are separated by 0.25 dex. The small circle denotes the location of the solar system. At low energies, the diffusion time scale and the time scale to escape the galaxy are comparable. As a consequence, it is possible to see the advection of the disk relative to the spiral arms. At higher energies, the escape is much faster and the advection cannot be seen.  }
\label{fig:GalaxyTop}
\end{figure*}

\begin{figure}[h]
\centerline{\includegraphics[width=3.4in]{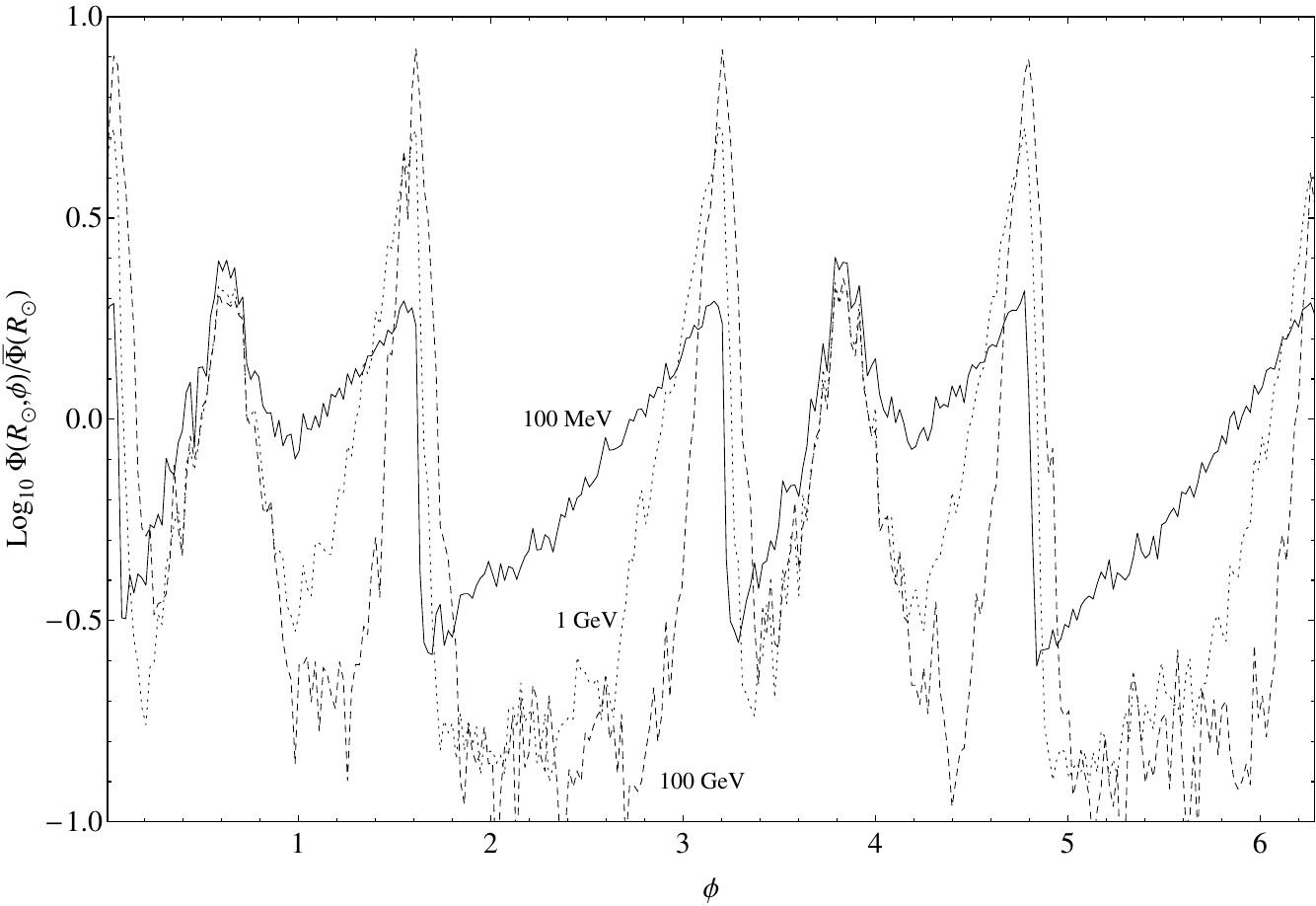}} 
\caption{The flux $\Phi(R_\odot,\phi$ at the galactic radius of the solar system as a function of the azimuthal angle, normalized to the average flux at this radius at the given energy. The solid line corresponds to the flux at 1 GeV, while the dashed to 100 GeV. Constraints based on Iron meteorites imply that the flux should vary by a factor of at least 2.5, between the spiral arms and the inter-arm region \citep{ShavivNewAstronomy}.}
\label{fig:azimuthal}
\end{figure}

After having validated our numerical simulations by recovering previous results for models with a homogeneous galacric source distribution, we can proceed to present the results of our model, where a significant fraction of the CRs is formed near the galactic spiral arms. 

When choosing model parameters, one has to remember that previously considered ``canonical" values are not binding. The reason is that the different path length distribution expected in the spiral arm model, which lacks short path lengths, requires a smaller typical diffusivity and halo size than the standard values obtained in the homogeneous models if the model is to fit observations \citep{Pamela}. Our goal in the present work is not to carry out an extensive parameter study but instead to show that model parameters exist for which the different isotopic spectra are recovered (see discussion in \S\ref{sec:discussion}).

The parameters of Model C that we use are $D = 5.5 \times 10^{26}$cm$^2$/sec, $Z_h=250$pc and $\delta = 0.5$. Following actual spiral arm passage about 20 Myr ago we assume that the peak CR acceleration took place 6 Myr ago. The 4-arms have a pitch angle of $i=28^\circ$, while the 2-arm set has $11^\circ$. Other parameters are as described in \S\ref{sec:model}.

Fig.\ \ref{fig:GalaxyTop} depicts the CR density at the galactic plane for two energies.  Clearly, the inhomogeneous source distribution gives rise to an inhomogeneous CR distribution in the galactic disk. 
Moreover, the effects of the different time scales is evident as well. At low energies, the spiral arm passage and escape time scales are comparable. This implies that the advection relative to the spiral arms becomes important, and one can see the ``smearing" associated with it. At high energies, the escape is much faster and one can consider the spiral arms to be effectively at rest (cf \S\ref{sec:BoCana}). 

\begin{figure}[ht]
\centerline{\includegraphics[width=3.4in]{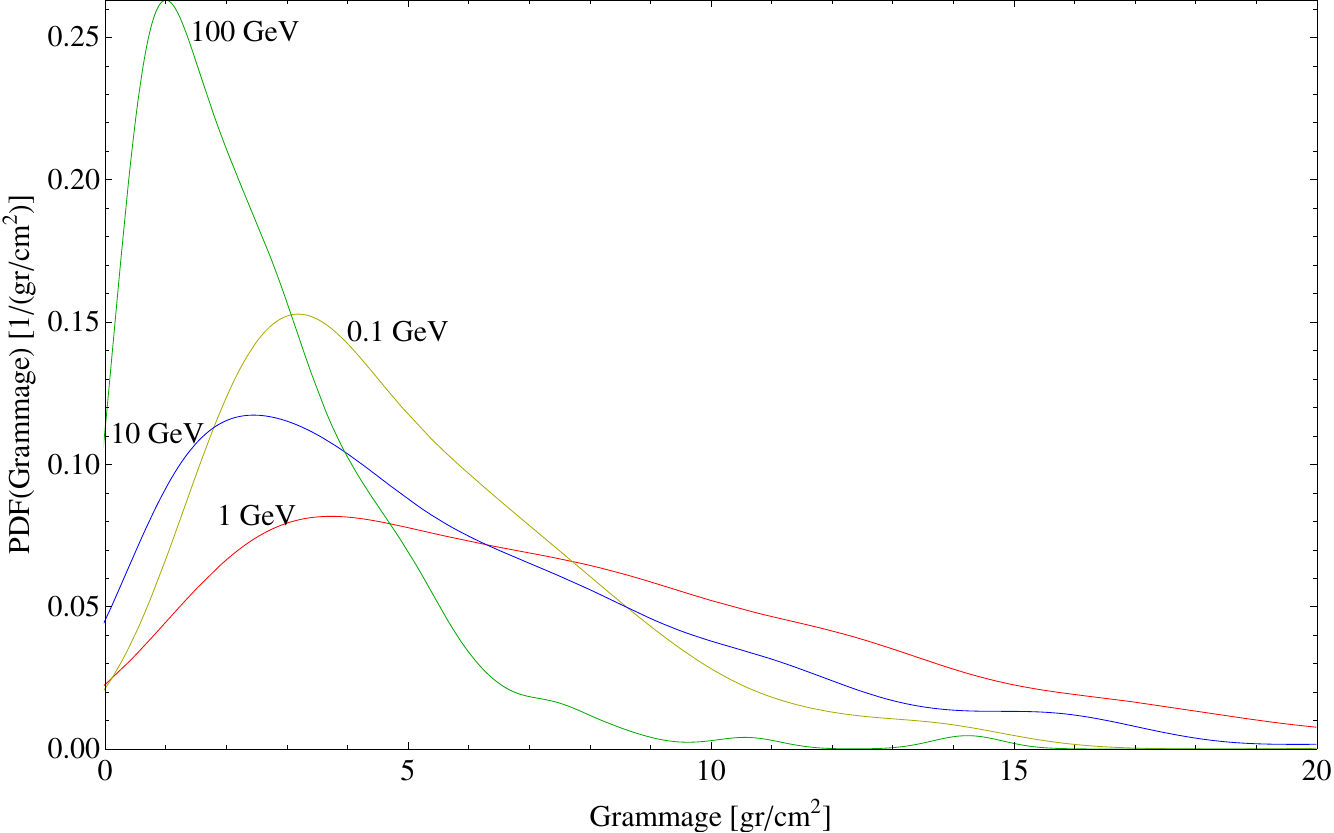}} 
\caption{The Path Length Distribution in our nominal model.
The energies are: 0.01GeV-yellow, 0.1GeV-purple, 1GeV-green and 10GeV-red. 
 Because most of the CRs need to travel from the nearby spiral arm, we find that there is a paucity in short path lengths. }
\label{fig:PLDours}
\end{figure}

The path length distributions that we obtain at several energies is depicted in fig.\ \ref{fig:PLDours}. One of the unique features of the present model is that the distributions are not exponential. Instead, they center around the typical grammage required for diffusion from the nearby arm. As we shall elaborate in the discussion, this non-exponential behavior has various interesting ramifications. The difference between the PLD in Model B (Fig.\ \ref{fig:PLDgalprop}) to the PLD in Model C ( Fig.\ \ref{fig:PLDours}) is the underlying reason why in our model we do not need a large halo.

Another interesting result is the different spectral indices of CR inside and outside the spiral arms. A non-dynamic model can give no spectral difference. Here, however, fig.\ \ref{fig:azimuthal} reveals that the ratio between the density at different energies becomes location dependent, with the ratio becoming ``harder" in the spiral arms. For example, the ratio between 100 GeV CRs and 1 GeV CRs in the wake of the arms can be as large as $10^{0.5}$ lower. In other words, the spectral index can soften by about 0.25 when leaving the arms. This is consistent with observations of $\gamma$-ray emission, which show that the spectrum in the direction of the orion arm has a spectral index harder by $0.4 \pm 0.2$ than in a high galactic latitude direction \citep{rogers,Bloemen}.

\subsection{B/C}

\begin{figure}[h]
\centerline{\includegraphics[height=2.0in]{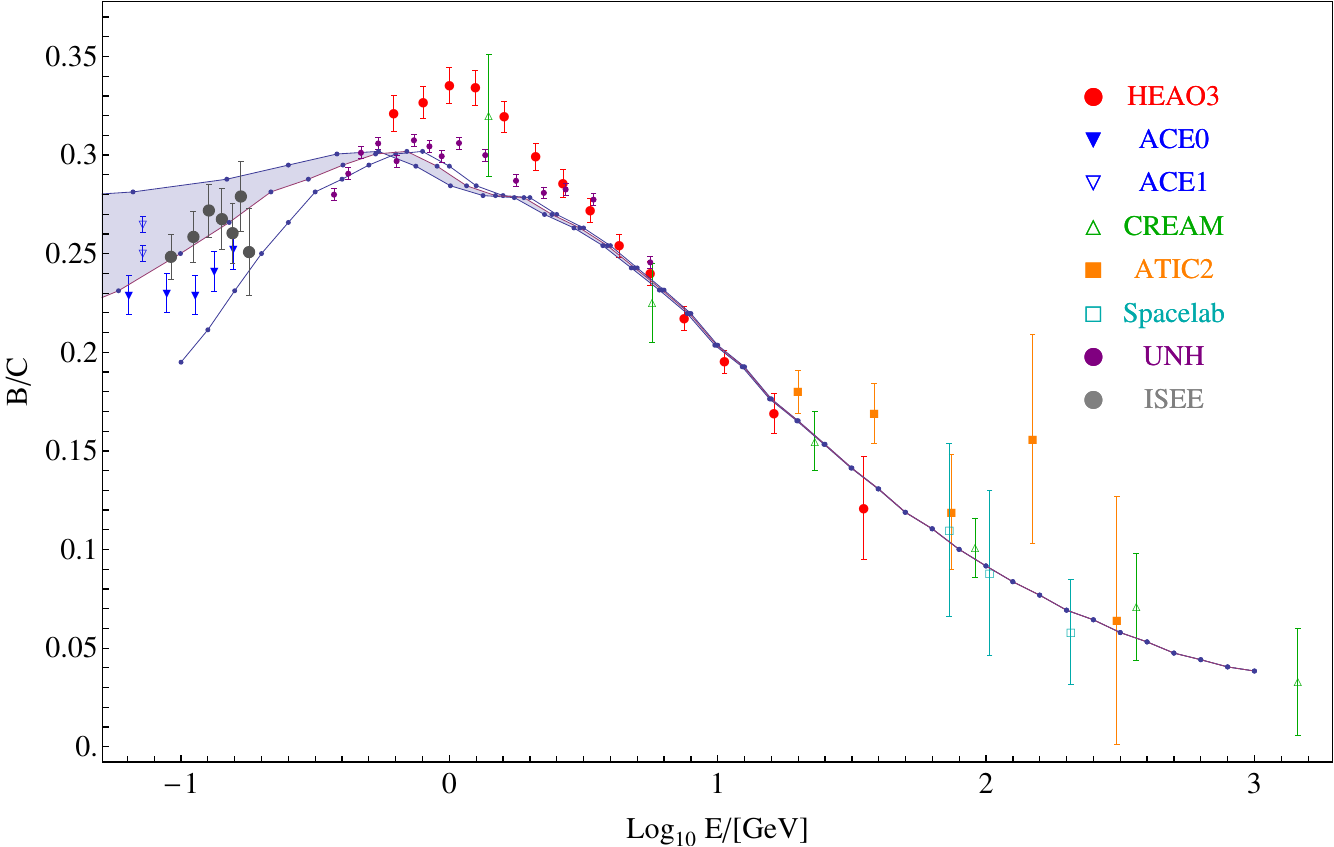}}
\caption{The Boron to Carbon ratio we obtain in our nominal model. }
\label{fig:BCarms}
\end{figure}

As predicted in \S\ref{sec:BoCana}, adding the dynamic spiral arms saturates the age at low energies, and gives a B/C ratio that increases with energy. Fig.\ \ref{fig:BCarms} depicts the B/C for our nominal model. 

\subsection{$^{10}$Be / $^{9}$Be}

\begin{figure}[h]
\centerline{\includegraphics[height=2.0in]{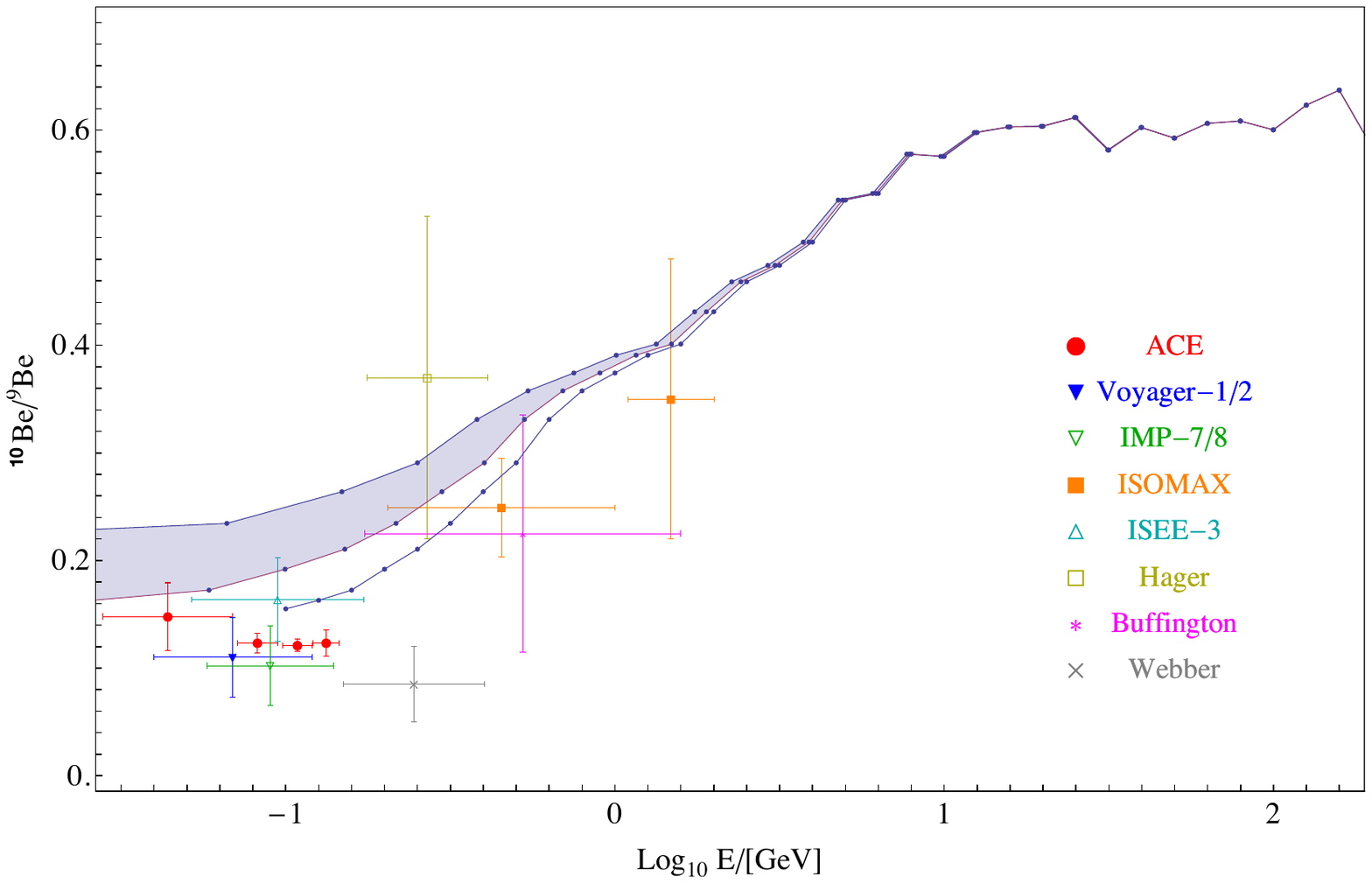}}
\caption{The $^{10}$Be to $^9$Be ratio we obtain in our nominal model. The results agree with the observation. The saturation below 1GeV is the result of the fixed age of the CR particles, which is caused by the dynamical arms.}
\label{fig:Be10Be9arms}
\end{figure}

The $^{10}$Be/$^{9}$Be provides information on the typical age of the CRs at different energies. The actual relation between the average age and the $^{10}$Be/$^{9}$Be ratio depends on the path length distribution. Therefore, we compare the model predictions with the $^{10}$Be/$^{9}$Be ratio directly. Fig.\ \ref{fig:Be10Be9arms} depicts the $^{10}$Be to $^9$Be ratio we obtain and its consistency with observations.

\subsection{Additional nuclear ratios}
\label{sec:CO}

\begin{figure}[h]
\centerline{\includegraphics[height=2.0in]{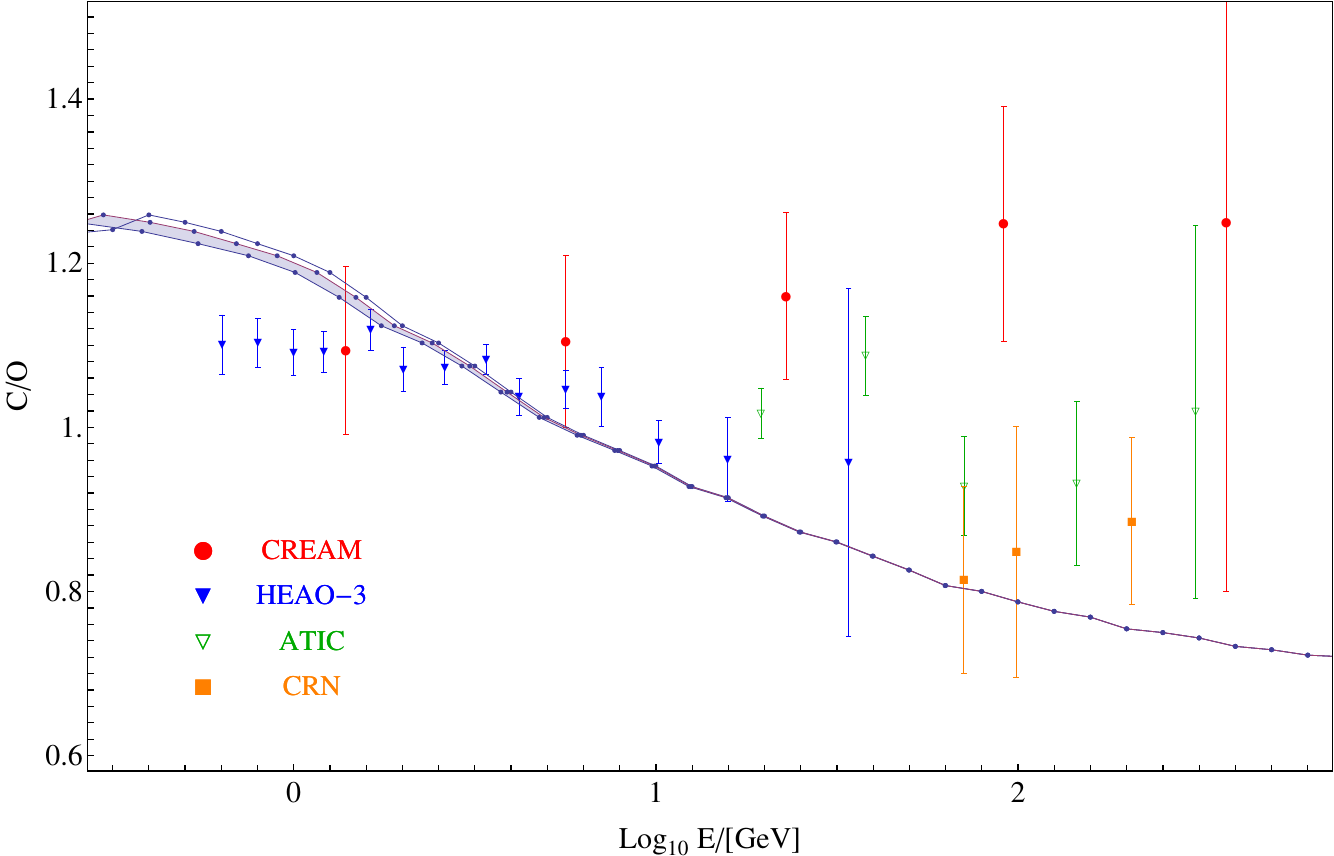}}
\caption{The Carbon to Oxygen ratio obtained in the model. The results agree with the observed energy dependance.}
\label{fig:COarms}
\end{figure}

To verify that the initial ratio at the sources of  C to O is consistent, we plot the Carbon to Oxygen ratio we obtain in fig.\ \ref{fig:COarms} and compare it with the observed ration. 

\begin{figure}[h]
\vskip 4mm
\centerline{\includegraphics[height=2.0in]{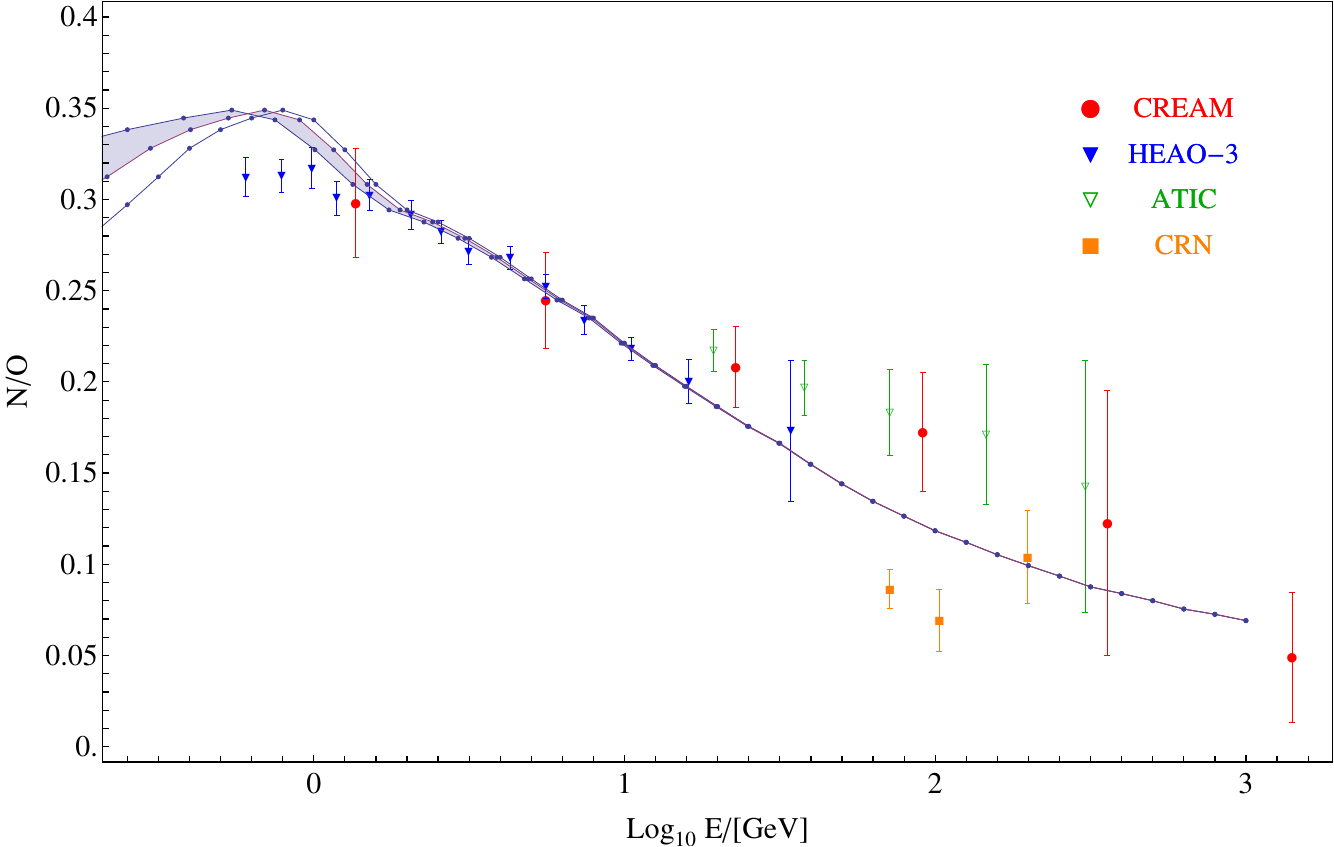}}
\caption{The N/O ratio. Since Nitrogen has a relatively large secondary component, its ratio to Oxygen should behave similar to the B/C ratio. }
\label{fig:NOarms}
\end{figure}

Because our model considers all isotopes up to $^{16}$O, we can also predict the Nitrogen to Oxygen ratio (note that Nitrogen in CRs is mostly secondary). The behavior is as expected similar to the B/C ratio. (See fig.\ \ref{fig:NOarms}.) However, it should be noted that the slope of the predicted C/O and N/O ratios is larger than the observed slope. Namely, there is some unexplained inconsistency between the power law index $\delta$ required to best fit the B/C ratio and the $\delta$ which best fits the $C/O$ and $N/O$ ratios. This inconsistency exists in the standard model and it remains here as well. 

\section{Discussion \& Summary}
\label{sec:discussion}

It is generally accepted that the bulk of the galactic cosmic rays (whether in number or in energy) are particles accelerated in supernova remnants. It is also generally accepted that most SNe in the Milky Way are core-collapse supernovae of massive stars, and that most of these stars are born and die near the galactic spiral arms \citep[e.g.,][and references therein]{ShavivNewAstronomy}. Yet, cosmic ray diffusion models generally do not take this into account. The exception is the work of \cite{ShavivNewAstronomy} who looked at the average CR reaching the solar system, and its variation due to the passage through the galactic spiral arms, and the work of \cite{Pamela} who considered the effect of the spiral arms on the positron to electron ratio. In the former work, it was shown using the exposure age of Iron meteorites, that the CR flux was probably variable, with the density in spiral arms at least 2.5 times larger than in between. In the latter work it was shown that one expects an in crease in the positron to electron ratio, as discovered by {\sc pamela} and confirmed by {\sc ams}, when taking the non-uniform distribution of CRs sources into consideration.

In this work, we developed the first fully 3D model which considers {\bf dynamic} spiral arms. This is essential when considering low energy particles that diffuse relatively slowly. We have shown that when coupling the spiral structure to the arm dynamics, one can recover the observed secondary to primary ratios (in particular, the Boron to Carbon ratio).
Instead of a monotonic decrease with energy, expected in the basic galactic diffusion model, one obtains a ratio that increases below 1GeV/nucleon and  decreases above this energy. This is because the age of the CRs at low energies is not determined by the diffusion time from the spiral arms, but instead by the time since the last spiral arm passage. This saturates the age. Since below 1GeV/nucleon, the particles are non-relativistic, a saturated age gives a grammage which is increasing with energy, and correspondingly an increasing secondary to primary ratio at low energies. 

This observed behavior of the B/C ratio at low energies has been known for almost 3 decades. Without any apparent physical mechanisms to explain it, several possibilities where devised. These include  a galactic cosmic ray wind, or re-acceleration or artificially having a diffusion coefficient that is constant below a few GeV/nucleon. We have shown that all these assumptions are not necessary (though at least two of them should be present at least at some level). Instead one simply needs to consider the theoretically expected and empirically observed source distribution. 

One very important aspect of this model is that the {\em path length distribution} is different from the one found in standard diffusion models. In the latter, the PLD is typically close to that of an exponent. However, if most CRs arrive from a distance, then the PLD will be missing the small path lengths (compare fig.\ \ref{fig:PLDgalprop} to fig.\ \ref{fig:PLDours}).  In this respect, it is very interesting to note that \cite{GarciaMunoz}  inferred from a  comparison of the B/C ratio to the sub-Fe/Fe ratio,  that the cosmic ray PLD should be missing short path lengths. This is because an exponential PLD would have given a smaller sub-Fe/Fe ratio for the grammage which explains the B/C.

The different PLD has another interesting ramification. CRs arriving after having passed a short path length necessarily stayed close to the galactic plane. In contrast, CRs having a long path length could stay further from the plane before returning back. Since the ISM density falls with the distance from the plane, CRs with short paths therefore experience a higher {\em average} ISM density than CRs with long paths. Thus, a distribution which is missing the short path lengths, as is the case in the spiral arm model, will have a lower average interaction with the ISM for a given average path length. In other words, the average grammage will be lower for the same average physical length. This result implies that if the spiral arm model is to recover the observed secondary to primary ratio, the model has to keep the CRs closer to the galactic plane where the density is higher. For this reason, the typical halo size required is lower (typically a few hundred pc, compared with the 1 to 4 kpc in more standard diffusion models). However, because the typical age is closely related to the ratio between the radioactive and stable Beryllium isotopes, which is a model independent observation, the smaller halo requires a lower diffusion coefficient. At 1 GeV/nucleon it should be of order $10^{27}$cm$^2/$s and not $10^{28}$cm$^2/$s.

This change in the derived diffusion coefficient and halo size elucidate that the canonical values for the parameters describing the cosmic ray diffusion were obtained under a given set of models. Once we change the models, we should reanalyze the various parameters accordingly, and not assume that their canonical values still hold.

Nevertheless, one has to confirm that the model predictions are consistent with the relevant observations, and in fact, there are several indications that a smaller halo and an inhomogeneous source distribution is required. When studying the synchrotron radiation from the edge on galaxy NGC 891 one finds a disk with a FWHM which is less than 500 pc \citep[as obtained by][for a distance of 10 Mpc]{Radio}. Although it could correspond to a slowly decaying electron density with a fast decaying magnetic field, a standard assumption of equipartition would require the cosmic ray density to decay fast with height. Although it is indicative, it does not prove the Milky Way necessarily has a thin halo as well, since the Milky Way's synchrotron emission can also be consistently fitted with a thick cosmic ray halo \citep{Orlando2013}. 
In $\gamma$-rays arising from $\pi$ decay, the observations are more direct. When fitting the Milky Way disk in  $\gamma$-rays using a standard homogeneous model with a large halo, one is left with a residual narrow disk having a half width of about 5$^\circ$ \citep{Residual}. Evidence for an inhomogeneous source distribution was obtained by FERMI. When fitting the energy dependent emissivity, one finds a harder spectral index in the Perseus and Local arms, by about 0.12, than the local spectrum \citep{Fermi}. Any time independent source distribution cannot explain this observation.

As a last note, it is very encouraging to see that the simple astrophysical source model with a realistic spatial distribution but with no new physics reproduces the observed B/C ratio, and in particular, its puzzling behavior at low energies. Even more encouraging is that the same inhomogeneous source distribution has the potential of explaining the puzzling increase in the positron fraction above $\sim$10 GeV \citep{Pamela}. If other observations, such as the apparent inconsistency between the sub-Fe/Fe ratio and B/C ratio \citep{GarciaMunoz} will also end up being explained by the model, then the circumstantial evidence will strongly imply that CR-source inhomogeneity plays an important role in CR physics.  Nevertheless, this change in the overall nature of the CR propagation requires revising some of the standard diffusion parameters, such as the halo size and diffusion constant, not because the observations change, but because their model dependent interpretation does.  
\def\jcap{J.\ Cos.\ Astropart.\ Phys.}
\def\na{New Astro.}

This work is supported by an Advanced ERC grant: GRB (DB and TP), by an ERC
starting grant and ISF grant no.\ 174/08 (EN), and by the I-CORE Program of the Planning and Budgeting Committee and The Israel Science Foundation (1829/12).

\bibliography{BoverC_withSpiralArms}

\end{document}